\documentclass[twocolumn,pra,superscriptaddress,nofootinbib,showpacs,10pt,aps]{revtex4-1}

\usepackage{amsmath}
\usepackage{amssymb}
\usepackage{graphicx}
\usepackage{color}

\usepackage[T1]{fontenc}
\usepackage[utf8]{inputenc}

\usepackage{amsthm}
\usepackage{bbm}
\usepackage{mathtools}

\usepackage{color}

\def\idty{{\bf 1}}

\let\eps\varepsilon

\newcommand{\spa}{E^{\mathbf a}}

\newcommand{\bor}{\mathbf{r}}
\newcommand{\boa}{\mathbf{a}}
\newcommand{\bob}{\mathbf{b}}

\newcommand{\boc}{\mathbf{c}}
\newcommand{\bod}{\mathbf{d}}

\newcommand{\boe}{\mathbf{e}}
\newcommand{\bof}{\mathbf{f}}

\newcommand{\tic}{\overline{C}}
\newcommand{\tid}{\overline{D}}

\newcommand{\bos}{\mathbf{s}}
\newcommand{\bok}{\mathbf{k}}
\newcommand{\boi}{\mathbf{i}}
\newcommand{\boj}{\mathbf{j}}

\newcommand{\bou}{\mathbf{u}}

\newcommand{\bosig}{{\boldsymbol\sigma}}
\newcommand{\inpr}[2]{{#1}\cdot{#2}}

\newcommand{\no}[1]{\|#1\|}

\newcommand{\h}[1]{\mathcal{#1}}

\newcommand{\R}{\mathbb{R}}  
\newcommand{\C}{\mathbb{C}}  

\newcommand{\stfm}{\mathfrak I}

\newcommand{\epno}{\eps_{\text{\sc no}}}
\newcommand{\etno}{\eta_{\text{\sc no}}}

\begin{document}
\title{Heisenberg uncertainty for qubit measurements}

\author{Paul Busch}
\email{paul.busch@york.ac.uk}
\affiliation{Department of Mathematics, University of York, York, United Kingdom}

\author{Pekka Lahti}
\email{pekka.lahti@utu.fi}
\affiliation{Turku Centre for Quantum Physics, Department of Physics and Astronomy, University of Turku, FI-20014 Turku, Finland}

\author{Reinhard F. Werner}
\email{reinhard.werner@itp.uni-hannover.de}
\affiliation{Institut f\"ur Theoretische Physik, Leibniz Universit\"at, Hannover, Germany}

\date{\today}
\begin{abstract}
Reports on experiments recently performed in Vienna [Erhard {\em et al}, {\em Nature Phys.} {\bf 8}, 185 (2012)] and Toronto  [Rozema {\em et al}, {\em Phys. Rev. Lett.} {\bf 109}, 100404 (2012)] include claims of a violation of Heisenberg's error-disturbance relation. In contrast, we have presented and proven a Heisenberg-type relation for joint measurements of position and momentum [{\em Phys. Rev. Lett.} {\bf 111},  160405 (2013)]. To resolve the apparent conflict, we formulate here a new general trade-off relation for errors in qubit measurements, using the same concepts as we did in the position-momentum case.  We show that the combined errors in an approximate  joint measurement of a pair of $\pm 1$-valued observables $A,B$ are tightly bounded from below by a quantity that measures the degree of incompatibility of $A$ and $B$. The claim of a violation of  Heisenberg  is shown to fail as it is based on unsuitable measures of error and disturbance. Finally we show how the experiments mentioned may directly be used to test our error inequality.   
\end{abstract}

\maketitle

\section{Introduction}
Heisenberg's error-disturbance relation \cite{Heisenberg1927} for measurements of incompatible quantities has recently become a popular subject of attack and proposed ``correction'' \cite{Erh12,Roz12,Baek13,Vienna2,Branciard2013,Weston-etal2013,diLor2013,Fuji2012}. Thus it appears as if one of the fundamental tenets of quantum mechanics is called into question; if the challenge proved tenable, it would have far-reaching ramifications for the status of the Heisenberg limit in precision measurements studied in the booming field of quantum-enhanced metrology \cite{Guta2012,BorSor2013}.  In contrast, in \cite{BLW2013c} we presented a Heisenberg-type error-disturbance relation for position and momentum. This result appears to contradict claims of an experimental  violation of Heisenberg's relation made in \cite{Erh12,Roz12,Baek13,Vienna2}. A direct comparison is made difficult by the fact that the experiments were performed on qubits rather than continuous variable systems.  Therefore, we will describe here the qubit variant of \cite{BLW2013c}.

The apparent conflict is then resolved by analyzing the meaning of the quantity, $\epno$, proposed by M. Ozawa (e.g., \cite{Ozawa04}), and adopted by the authors of \cite{Erh12,Roz12,Baek13,Vienna2} and others. This quantity  is defined suggestively  as the square root of the expectation of a squared {\em noise operator}. However, we will see that it does {\em not} meet its intended purpose of  representing state-specific experimental errors but something else. Therefore $\epno$ provides {\em no basis for claims of a theoretical or experimental violation of Heisenberg-type error-disturbance relations.} Actually, the experiments confirm Ozawa's inequality and demonstrate a violation of the (incorrect) inequality  $\epno(A,\rho)\epno(B,\rho)\ge|\langle[A,B]\rangle_\rho|$, which is attributed wrongly  to Heisenberg  (who never gave a  quantum mechanical definition of measurement errors or proposed a precise inequality of this generality). 

 In contrast,  our approach represents measurement error as an overall figure of merit of the measuring device, giving a worst-case estimate of the inaccuracy applicable to all possible input states. Our error measure $\Delta$, introduced in \cite{BLW2013c}, is an {\em operationally significant} quantum version of the classic root-mean-square error, obtained by an adaptation of the so-called Wasserstein distance (of order 2) between probability distributions \cite{Villani}. It can be applied seamlessly to the qubit case, yielding our main result, a Heisenberg-type error uncertainty relation (Sec.~\ref{sec:tradeoff}): {\em any joint measurement of two-outcome observables $C,D$ has combined approximation errors that are constrained by a measure of the degree of incompatibility of the target observables $A,B$ to be approximated.} Symbolically:
 \[
 \Delta\bigr(C,A\bigr)^2+\Delta\bigr(D,B\bigr)^2 \ \ge \ (\text{incompatibility of $A,B$}).
 \]

The additive form of this trade-off relations offers itself given that an error product cannot have a nonzero bound. This raises the question of whether the traditional uncertainty relation for the spreads of two observables in a quantum state can be supplemented with an additive version. We answer this in the positive (Sec.~\ref{sec:addit}), with an inequality for the sum of the variances of  $A,B$ in state $\rho$,
\[
\Delta(A,\rho)^2+\Delta(B,\rho)^2\ \ge \ (\text{noncommutativity of $A,B$}),
\]
where the bound is state independent and is nontrivial also for eigenstates of $A$ or $B$.

The proofs of these inequalities are based on simple geometric considerations, which makes it possible to teach them in  a basic quantum mechanics course.  

Ironically, Ozawa's measure $\epno$ is actually state independent in the class of qubit measurements under consideration here and thus over-estimates badly the state-dependent errors (Sec.~\ref{sec:Oz}). In fact, \textsl{rather than helping to prove Heisenberg wrong, the quantity $\epno$ itself satisfies a Heisenberg-type trade-off inequality}, with the same bound as for our quadratic error inequality:
\[
\epno(A,\rho)+\epno(B,\rho)\ \ge \ \tfrac 12(\text{incompatibility of $A,B$}).
\]
(For a more general, detailed critique of the noise-operator based approach of attempting to quantify measurement errors, we refer the reader to our forthcoming investigation \cite{BLW2013a}; there we also amplify on the fact that it is historically incorrect to associate Heisenberg with the above wrong inequality.)

We finally (Sec.~\ref{sec:expt}) proceed to demonstrate the possibility of using the setups of the Vienna and Toronto  experiments to  test our qubit measurement error relation. The Toronto experiment  allows the realization of the tight error bound. Importantly, the experiments and their analyses reported so far are in fact \textsl{incomplete}: they investigate and confirm a mathematical relation -- Ozawa's inequality -- between two quantum mechanical expectation values, $\epno(A,\rho)$ and $\epno(B,\rho)$, and this result is accompanied with the statement that hence the incorrect error-disturbance relationship attributed to Heisenberg has been tested and violated. There is no independent evaluation of the claim that these quantities do represent approximation errors; this assertion is adopted on faith from Ozawa. What is required for a test of measurement error trade-off relations is an error analysis in which the actual measurement statistics are compared to those of (more) precise reference measurements.

\section{A trade-off relation for quantum RMS errors}\label{sec:tradeoff}

We will consider a pair of sharp qubit observables $A,B$. 
Our aim is to characterize positive operator
valued measurements (observables) $C$, $D$ which are {\em compatible}, that is, they can be performed simultaneously, and which will be considered as approximations to $A$ and $B$, respectively. (We recall that two observables are compatible or jointly measurable if there is another, joint, observable of which they are marginals.)

\begin{figure}[ht]
\centering
  \includegraphics[width=8cm]{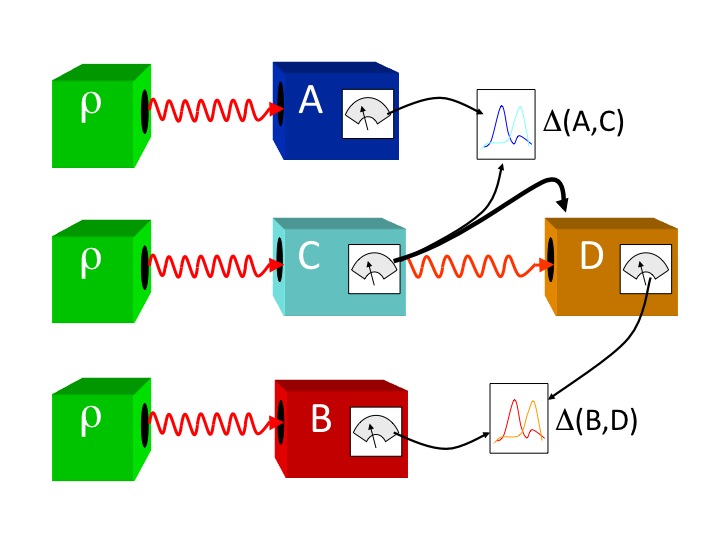}
	\caption{Sequence of compatible measurements $C$ and $D$. The statistics are compared with control measurements $A$ and $B$, respectively,
	defining the approximation errors.}
		\label{fig:seq}
\end{figure}

\begin{figure}[ht]
\centering
  \includegraphics[width=8cm]{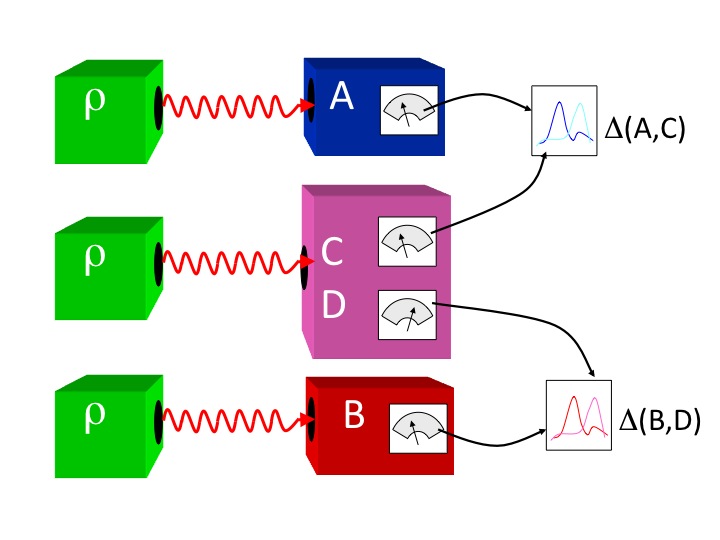}
	\caption{Scheme of a joint measurement of compatible observables $C,D$, each being used as an approximation of $A,B$, respectively.}
		\label{fig:joint}
\end{figure}

The problem of measurement disturbance and simultaneous approximation is illustrated in figures \ref{fig:seq} and \ref{fig:joint}. A measurement $C$ as an approximation to $A$ makes itself felt by changing the state of the system, so that a subsequent measurement of an observable $D$ will be an approximation of  $B$ only with limited accuracy if $A$ and $B$ are not compatible (figure \ref{fig:seq}). Such a scheme is a special case of a device in which the  boxes $C$ and $D$ are merged, giving a truly ``joint'' measurement 
(figure \ref{fig:joint}). Thus, a measure of disturbance is conceptually an instance of an approximation error. 

We use Bloch sphere notation to write the spectral projections of $A$ as $A_\pm=\spa_\pm=\frac12(\idty\pm\inpr\boa\bosig)$, and similarly for $B$, where $\boa,\bob$ are unit vectors. To be specific, we scale these measurements such that their outcomes are $\pm 1$, so that, for example, the observable $C$ is given as a map $\pm 1\mapsto C_\pm$, with the positive operators $C_+=\frac12(c_0\idty+\inpr\boc\bosig)$, $C_-=\idty-C_+$.  (Positivity is equivalent to  $\no{\boc}\le\min\{c_0,2-c_0 \}\le1$.) 

The first task is to specify an error measure to quantify the quality of such approximations. We follow the choice made in \cite{BLW2013c}. For a pair of observables $E:\pm1\mapsto E_\pm$ with $E_+=\frac12(e_0\idty+\inpr\boe\bosig)$ and $F:\pm\mapsto F_\pm$,  $F_+=\frac12(f_0\idty+\inpr\bof\bosig)$, a state-dependent distance that scales with the values and quantifies the difference between the probability distributions $E_\rho$ and $F_\rho$ (where $\rho$ denotes the state) is given by the Wasserstein distance:
\[
\Delta\bigr(E_\rho,F_\rho\bigr)=\left[\inf_\gamma \iint (x-x')^2d\gamma(x,x')\right]^{\frac12},
\] 
where the infimum is taken over all {\em couplings} $\gamma $ of $E_\rho,F_\rho$ (i.e., all joint distributions with marginals $E_\rho,F_\rho$). By maximizing the state-dependent error over all states $\rho$ one has the worst-case error estimate,
$
\Delta\bigr(E,F\bigr)=\sup_\rho \Delta\bigr(E_\rho,F_\rho\bigr)
$.

In the present case of $\pm 1$-valued qubit observables it is straightforward to write down all possible couplings and determine this infimum (Appendix A). Writing the states as $\rho=\frac12(\idty+\inpr\bor\bosig)$, one obtains
$$
\Delta\bigr(E_\rho,F_\rho\bigr)^2=2\bigl |e_0-f_0+\inpr\bor{(\boe-\bof)}\bigr| .
$$
By maximizing this over all states $\rho$ one has the worst-case error estimate
\[
\Delta\bigr(E,F\bigr)^2=2|e_0-f_0|+2\no{\boe-\bof}.
\]

We pause to emphasize that one could use the state-dependent error measure for the study of precision measurements in which one is interested in an error-disturbance trade-off in a specific state. However, for an assessment of the quality of a joint measurement device, it is also important to note that one can always arrange for situations where the state-dependent errors are both zero: for example, one can take $C$ and $D$ both sharp and identical, with $\boc=\bod$ in the plane spanned by $\boa,\bob$. Then for a state $\rho$ with $\bor$ perpendicular to that plane one has $\Delta(A_\rho,C_\rho)=\Delta(B_\rho,D_\rho)=0$.

We could also have chosen different distance measures for the comparison of two observables.
For  comparison we note the distance induced by the total variation norm (also known as 1-norm), which was used in \cite{BuHe08} for a
similar purpose and turns out to be given as $D\bigr(E,F\bigr)=\frac 14 \Delta\bigr(E,F\bigr)^2$ in this special qubit situation.

The Heisenberg-type joint measurement error trade-off relation that we present here gives a tight lower bound for the sum of the squared approximation errors:
for any pair of observables $C$ and $D$ that are jointly measurable, their errors of approximation relative to $A$ and $B$ are tightly bounded as follows.
\begin{align}\label{QUR}
%\begin{split}
\Delta\bigr(C,A\bigr)^2&+\Delta\bigr(D,B\bigr)^2 %\\
\ge \sqrt{2}\bigl[\no{\boa-\bob}+\no{\boa+\bob}-2  \bigr]\nonumber\\
&\quad =\frac1{\sqrt2}\bigl[ \Delta\bigr(A,B\bigr)^2+\Delta\bigr(A,B^{(-)}\bigr)^2-4 \bigr].
%\end{split}
\end{align}
Here $B^{(-)}$ is the observable obtained from $B$ by swapping the outcomes $\pm 1$. We will see that the lower bound represents the incompatibility of $A$ and $B$.

The proof procedure (which we sketch in Appendix B)  follows the same steps as that of the position-momentum case \cite{BLW2013c}. First one reduces the inequality to the special case where the estimating observables are {\em covariant} under value translations (swaps). For such estimators we can then directly give a simple geometric proof of the tight bound. We find this bound by minimizing the left hand side of \eqref{QUR} under the constraint of compatibility of the covariant approximators $\tic_\pm=\frac12(\idty+\inpr\boc\bosig)$, $\tid_\pm=\frac12(\idty+\inpr\bod\bosig)$, a criterion of which is given by the following inequality  \cite{Busch86} (see also Appendix C):
\begin{equation}\label{jm1}
\no{\boc-\bod}+\no{\boc+\bod} 
\le 2.
\end{equation}
For an incompatible pair of  observables $A,B$, it is thus natural to define their {\em degree of incompatibility} by the (positive) number
$\no{\boa-\bob}+\no{\boa+\bob}-2$.

According to \eqref{jm1}, compatibility does not require commutativity. It is only when at least one of the observables is sharp (projection valued)
 that compatibility implies commutativity. The lower bound in \eqref{QUR} 
reaches its minimal value zero exactly when the sharp observables $A,B$ are compatible. (For a pair of compatible unsharp observables this number
can be negative.) The  bound 
reaches its maximum $2(2-\sqrt2)$ when $\boa,\bob$ are orthogonal unit vectors; then equality in \eqref{QUR} is achieved for the covariant
observables $\tic,\tid$ with $\boc=\boa/\sqrt2$ and $\bod=\bob/\sqrt2$. 
In all other cases, the optimal approximations are obtained for vectors $\boc,\bod$ that are not collinear with $\boa,\bob$.

We will show in Section \ref{sec:prep-ur-mmt-ur} below that the covariant estimators $\tic,\tid$ are smearings of their sharp counterparts, and analysis of the smearing operation shows that the error trade-off can be reduced to a form of {\em preparation} uncertainty relation. This observation, which was also made in the position-momentum case, corroborates an intuition held by the pioneers of quantum mechanics: {\em the possibilities of measurement cannot exceed the possibilities of preparation.}

We thus see how inequality \eqref{QUR} limits the combined approximation accuracies, and the tight bound is given by the 
incompatibility degree of the target observables $A,B$ being approximated. The minimum is taken under the constraint of joint measurability
of the approximating observables. The incompatibility degree 
is determined by the average of the squared distances between $A,B$ and $A,B^{(-)}$, respectively, reflecting the fact that compatibility is independent of the choice of scaling of the outcomes.

We can also give a somewhat different error trade-off relation that is closer in form and spirit to the position-momentum inequality of \cite{BLW2013c}.
Note that one has
\begin{align*}
\Delta(C,A)^2&=2\bigl[\,|c_0-1|+\no{\boc-\boa}\,\bigr]\\
&\ge  2(1-\no{\boc})\ge1-\no{\boc}^2\equiv U(\tic)^2,
\end{align*}
and similarly $\Delta(D,B)^2\ge U(\tid)^2$, where $U(\tic)$ is a measure of the degree of {\em unsharpness} of the covariant observable
$\tic$, that is, its deviation from being projection valued. A simple calculation \cite{BuHe08} shows that the compatibility condition \eqref{jm1} is 
equivalent to
\begin{equation}\label{jm2}
U(\tic)^2U(\tid)^2\ge \no{\boc\times\bod}^2=4\no{[\tic_+,\tid_+]}^2.
\end{equation}
This inequality says that two noncommuting (covariant) observables are compatible if and only if they are sufficiently unsharp. Sharpness of one of them
forces commutativity.
Thus we also obtain a bound for the error product:
\begin{equation}\label{QUR2}
\Delta(\tic,A)^2\,\Delta(\tid,B)^2\ge 4\no{[\tic_+,\tid_+]}^2.
\end{equation}
Here we see how the noncommutativity of the {\em compatible} approximators limits the accuracies. However, one may
choose to approximate $A,B$ using commuting observables $C,D$. In this case the approximation will not be optimal but the bound for the error product vanishes. This highlights the relative strength of the bound \eqref{QUR} for the sum of squared errors.

\section{Products or sums of uncertainties?}\label{sec:addit}

It has become accepted wisdom that uncertainty relations have the form of a lower bound for an uncertainty {\em product}. In contrast,  \eqref{QUR} gives a lower bound to a {\em sum} of uncertainties. From the discussion above it is
evident that there is no nontrivial lower bound for the product of errors.  To help appreciate this less conventional perspective, we note here an {\em additive} version of a preparation uncertainty relation. The standard deviation of a sharp $\pm1$-valued qubit observable $A$  in a state $\rho$ is $\Delta(A,\rho)=\bigl[1-(\inpr\bor\boa)^2\bigr]^{1/2}$. Then
\begin{equation}\label{sum1}
\Delta(A,\rho)+\Delta(B,\rho)\ \ge\ \no{\boa\times\bob}=2\no{[A_+,B_+]}.
\end{equation}
The left hand side is equal to $\no{\bor\times\boa}+\no{\bor\times\bob}$, so one can see that the bound is attained for $\bor=\pm\boa$ or
$\bor=\pm\bob$.

We can also minimize the sum of the variances:
\begin{align}\label{sum2}
\Delta(A,\rho)^2+\Delta(B,\rho)^2&\ge 1-|\inpr\boa\bob|%\nonumber\\
%&\quad
=1-\sqrt{1-\no{\boa\times\bob}^2}\quad\nonumber\\
&=1-\sqrt{1-4\no{[A_+,B_+]}^2}.
\end{align}
Again, the bound is tight, but this time it is attained at $\bor=(\boa\pm\bob)/\no{\boa\pm\bob}$ for $\inpr\boa\bob\ge 0$ and $\le 0$, respectively.

Inequalities \eqref{sum1}, \eqref{sum2} are stricter than the state dependent bound  for the product of the standard deviations: here we obtain a
nontrivial lower bound also when $\rho$ is an eigenstate of $A$ or $B$. The lower bounds in both \eqref{sum1} and \eqref{sum2} vanishes
exactly when $A$ and $B$ commute.

It is interesting to compare this situation with the case of position $Q$ and momentum $P$. 
Let $x_0$ be an arbitrary positive constant of the dimension of length. It is an easy exercise
to show that for any value of $x_0$ the inequality
\begin{equation}\label{sum3}
\frac{4\hbar^2}{x_0^2}\Delta(Q,\rho)^2+x_0^2\Delta(P,\rho)^2\ge 2\hbar^2
\end{equation}
is a consequence of 
\begin{equation}\label{prod-UR}
\Delta(Q,\rho)\Delta(P,\rho)\ge\hbar/2.
\end{equation} 
Conversely, using the reciprocal behavior of position and momentum under scale transformations, it can be shown that if  inequality \eqref{sum3} is assumed to holds for only one value of $x_0$ and all
states $\rho$, then it holds for all values of $x_0$ and entails the standard uncertainty relation \eqref{prod-UR}.
Inequality \eqref{sum3} can also be proven directly (i.e., without making use of \eqref{prod-UR}) by observing that 
finding the minimum of the left hand side is equivalent to finding the minimum energy eigenstate of the harmonic oscillator Hamiltonian.

\section{Error bounded by uncertainty}\label{sec:prep-ur-mmt-ur}

There is a general connection between the limitations of preparations and the limitations of measurement: the possibilities of measurement should not exceed the possibilities of preparation; hence a limitation of the latter should entail a limitation of the former. We can see this principle at work in the present case of qubit measurements, in much the same way as it played a role in the case of position and momentum \cite{BLW2013c}.

We consider the case $\boa\perp\bob$.
If the approximator $C$ is a smearing of $A$, so that $C_+=\frac12(\idty+\lambda\inpr\boa\bosig)=\mu_+ A_++\mu_-A_-$ for a probability distribution $\mu$ with $\mu_++\mu_-=1$, then we find $\Delta\bigl(A,C \bigr)^2=2(1-\lambda)=4\mu_-\ge\Delta(\mu)^2$, since $\lambda=\mu_+-\mu_-$. Similarly we get $D_+=\frac12(\idty+\lambda\inpr\bob\bosig)=\nu_+B_++\nu_-B_-$, and $\Delta\bigl(B,D \bigr)^2=2(1-\lambda)=4\nu_-\ge\Delta(\nu)^2$. Now we observe that we can identify the distributions $\mu$ and $\nu$ with distributions of $A$ and $B$ with one and the same quantum state $\rho_\bos$, with $\bos=\lambda(\boa+\bob)$, $\lambda=\bos\cdot\boa=\bos\cdot\bob$:
\[
\mu_-=\tfrac12(1-\inpr\bos\boa)=\tfrac12(1-\inpr\bos\bob)=\nu_-,
\]
so that  $\Delta(\mu)^2=\Delta(A,\rho_\bos)^2\le \Delta\bigl(A,C \bigr)^2$ and $\Delta(\nu)^2=\Delta(B,\rho_\bos)^2\le \Delta\bigl(B,D \bigr)^2$.
Taking $\lambda=1/\sqrt2$, the largest value allowed by the compatibility of $C,D$, and using \eqref{sum2} we get:
\begin{align*}
\Delta\bigl(A,C \bigr)^2+\Delta(B,D)^2&=4\mu_-+4\nu_-=2(2-\sqrt2)\\
&\ge \Delta(A,\rho_\bos)^2+\Delta(B,\rho_\bos)^2\ge 1.
\end{align*}
Thus, if one did not know already that $2(2-\sqrt2)$ is the optimal bound for the combined squared errors, the uncertainty relation for the state $\rho_\bos$ would guarantee a bound. Moreover, the tight bound, given above in the form $4(\mu_-+\nu_-)$,  is itself a characteristic of the state operator $\rho_\bos$.

The role of the operator $\rho_\bos$ becomes more transparent by constructing a joint observable for the approximators $C,D$.  A general expression is given in Appendix C; it is easy to see that if $C,D$ (with $\boc\perp\bod$ and $c=\no\boc=\no\bod=d$) are compatible (that is, $c=d\le 1/\sqrt2$), then the following is a joint observable:
\begin{align*}
G_{+\pm}&=\tfrac14\bigl[\idty+(\boc\pm\bod)\cdot\bosig\bigr],\\ 
G_{-\pm}&=\tfrac14\bigl[\idty-(\boc\mp\bod)\cdot\bosig\bigr].
\end{align*}
We specify Cartesian coordinates with orthogonal unit vectors  $\boe_1,\boe_2,\boe_3$, such that $\boc=c\,\boe_1$, $\bod=d\,\boe_3$. Then $(k,\ell)\mapsto G_{k\ell}$ is covariant under the unitary group acting on operators, with elements 
\begin{align*}
\mathcal{U}_{++}&=\idty(\cdot)\idty,\quad\quad \mathcal{U}_{--}=\sigma_2(\cdot)\sigma_2,\\
\mathcal{U}_{+-}&=\sigma_2(\cdot)\sigma_2,\quad \mathcal{U}_{-+}=\sigma_3(\cdot)\sigma_3,
\end{align*}
where $\sigma_1,\sigma_2,\sigma_3$ are the Pauli operators associated with coordinate axes $x,y,z$.
This group can be cast as a representation of a discrete Heisenberg-Weyl group, and it is straightforward to verify that the joint observable can be given in the form
\[
G_{k\ell}=\tfrac12\mathcal{U}_{k\ell}(\rho_\bos),\quad\rho_\bos=\tfrac12\bigl[\idty+c(\boe_1+\boe_3)\cdot\bosig\bigr].
\]
This explains why the approximation errors in such a covariant measurement are determined by the uncertainties inherent in the state operator $\rho_\bos$.
Further discussion of error trade-off relations for discrete Heisenberg-Weyl covariant observables and their mutually unbiased marginals can be found in \cite{BLW2013b}.

\section{Interpretation of the noise-operator based measures}\label{sec:Oz}

The Vienna and Toronto experiments make use of covariant observables as approximators, and it turns out that the ``disturbed'' observables are covariant as well. The disturbance measure $\etno$ used there is in fact a variant of $\epno$, so that we can use unified notation. For  a covariant approximator $\tic$ of $A$ one obtains (we are using the notation $A[x^n]=\int x^ndA(x)$ for the $n^{\rm th}$ moment operator of an observable $A$):
\begin{align*}
&\epno(A,\rho)^2={\rm tr}\bigl[\rho(\tic[x^2]-\tic[x]^2)\bigr]
%&\qquad\qquad
+{\rm tr}\bigl[\rho(\tic[x]-A[x])^2\bigr]\\
&\qquad=1-\no{\boc}^2\ +\ \no{\boc-\boa}^2
=U(\tic)^2+\tfrac1{4}\Delta(\tic,A)^4.
\end{align*}
Here we see that $\epno$   is a mix of an error contribution and the intrinsic unsharpness of the estimator observable -- which is already accounted for in the $\Delta$ term; it is not hard to see that $\epno(A,\rho)\le \Delta(\tic,A)$. For approximators that are smearings of the target observable, for which $\boc=\gamma \boa$, one has in fact $\epno(A,\rho)=\Delta(\tic,A)$. This situation arises in the Toronto experiment (see below).

What is most striking is that in this particular case of covariant qubit observables, $\epno$ has lost what the advocates of this measure consider to be one of its virtues: its state-dependence. Thus $\epno$ is a bad overestimate of the state-dependent error; in particular, it cannot capture the peculiar situation arising in both the Vienna and Toronto experiments where the input and output distributions are identical, so that the state-dependent error vanishes.

The inequalities \eqref{QUR} and \eqref{jm2} immediately yield similar trade-off relations for the $\epno$ quantities.
In fact, using $\epno(A,\rho)\ge \frac 12 \Delta(\tic,A)^2$ (and similarly for $\epno(B,\rho)$, then \eqref{QUR} gives
\begin{align*}%\label{QUR-epno}
\epno(A,\rho)+\epno(B,\rho)
\ge \frac 1{\sqrt{2}}\bigl[\no{\boa-\bob}+\no{\boa+\bob}-2  \bigr];
\end{align*}
and using $\epno(A,\rho)\ge U(\tic)$, then \eqref{jm2} entails
\begin{equation*}%\label{jm-epno}
\epno(A,\rho)^2\epno(B,\rho)^2\ge \no{\boc\times\bod}^2=4\no{[\tic_+,\tid_+]}^2.
\end{equation*}
Thus, not surprisingly, the quantities $\epno$, which comprise a mix of contributions from error and unsharpness,
are seen to be subject to Heisenberg-type trade-off constraints.

As argued in \cite{BuHeLa04,Werner04} and elaborated further in \cite{BLW2013a}, $\epno$ is a problematic generalization of Gauss's root-mean-square error into the quantum context.  This quantity does not, in general, provide an operationally significant estimate of measurement errors in a single state.  One can see this already from the general defining expression for $\epno$ given above: the operator $\tic[x]$ does not, in general, commute with $A[x]$, so that the difference $\tic[x]-A[x]$ is in fact incompatible with both. Therefore it is not evident that a comparison of the statistics of the observables $A$ and $C$ can be obtained from studying their difference operator. This apparent deficiency has been addressed with the observation that $\epno^2$ can be expressed as a combination of expectation values of first or second moments of the approximator observable $C$ in {\em three different states} instead of just one. Accordingly, in the Vienna experiment the quantity $\epno$ is measured using the so-called three-state method. The fact that three distinct states are required makes evident the impossibility of interpreting this quantity as the error relevant to a {\em single} state. This is illustrated in the present qubit case by the above expression for $\epno$, which shows it to be state-independent and in fact related to our maximized error.

In higher dimensional Hilbert spaces it is not hard to construct examples of measurements where $\epno$ vanishes although the input and output distributions to be compared are not identical. There are also examples where these distributions do coincide but the quantity $\epno$ can be made arbitrarily large. Similar observations apply to the use of this quantity as a measure of disturbance, showing that these quantities are unreliable as indicators of error or disturbance \cite{BLW2013a}.

\section{Proposed experimental tests}\label{sec:expt}

We consider first an experiment of the kind performed in Vienna, where a projective (or von Neumann-L\"uders) measurement 
of a sharp observable $C$ (with $\no{\boc}=1$) is considered as an approximate measurement of $A$. Such a measurement causes the state change
$
\rho\to C_+\rho C_++C_-\rho C_-,
$
or equivalently, distorts an observable $B$ into $D$ as
\begin{align*}
B_\pm\to D_\pm \
&=  C_+B_\pm C_++C_-B_\pm C_- \\
&= B_{\rho_{\boc}}(\pm1)C_++B_{\rho_{\boc}}(\mp1)C_-\\
&=\tfrac12(\idty\pm\inpr\bod\bosig),\quad
\bod=(\inpr\boc\bob)\boc.
\end{align*}
(Here $\rho_\boc$ denotes a pure state with unit Bloch vector $\boc$, and $B_{\rho_{\boc}}(\pm1)={\rm tr}[\rho_\boc B_\pm]=\frac 12(1\pm\inpr\boc\bob)$.)
This scheme defines a joint observable $M$ (necessarily of the product form since $C$ is sharp), with 
positive operators
$
M_{k,\ell}= C_kD_\ell, \quad k,\ell=\pm
$, 
which can be considered as an approximate joint measurement of $A$ and $B$, with the characteristic errors
$\Delta(C,A)$ and $\Delta(D,B)$. 

The (squared) state-dependent error and disturbance are given by ($\rho=\frac 12(\idty+\inpr\bor\bosig)$)
\begin{align*}
\Delta(C_\rho,A_\rho)^2=2|\inpr\bor{(\boc-\boa)}|,\
\Delta(D_\rho,B_\rho)^2=2|\inpr\bor{(\bod-\bob)}|\,.
\end{align*}
We observe that if $\inpr\bor\boa=\inpr\bor\boc$, then $C_\rho=A_\rho$, so that the 
state-dependent $\Delta(C_\rho,A_\rho)=0$ in this case. 
Since $\Delta(D_\rho,B_\rho)\leq 2$,
the state dependent uncertainty product $\Delta(C_\rho,A_\rho)\Delta(D_\rho,B_\rho)=0$ for all such states.

The maximized error and disturbance are
\begin{align*}
\Delta(C,A)^2&=2\no{\boc-\boa}=2\sqrt2\sqrt{1-\inpr\boc\boa}\,,\\
\Delta(D,B)^2&=2\no{\bod-\bob}=2\no{\bob\times\boc}.
\end{align*}
These are nonzero if $\boc\ne\boa$ and $\bob\ne\boc$, respectively.

It is straightforward to show  that the following uncertainty relation holds for this experiment:
\begin{equation}\label{qubit-EDR1}
\begin{split}
\Delta(C,A)^2&+\Delta(D,B)^2
=2\no{\boc-\boa}+2\no{\bob\times \boc}\\
&\ge 2\no{\boa\times \bob}
=4\bigl\|{[A_+,B_+]}\bigr\|.
\end{split}
\end{equation}
The minimum is achieved for $\boc=\boa$. 

This kind of sharp measurement as an approximate joint measurement is not an optimal joint approximation: for example, in 
the case of orthogonal $\boa,\boc$, the lower bound is $2>2(2-\sqrt2)$. 

For comparison we give the squared quantities $\epno$:
\begin{align*}
\epno(A,\rho)^2 &=  {\rm tr}\bigl[\rho(C[x]-A[x])^2\bigr]= \no{\boc-\boa}^2 \\
\epno(B,\rho)^2  &= {\rm tr}\bigl[\rho(D[x^2]-D[x]^2) \bigr] +
{\rm tr}\bigl[\rho(D[x]-B[x])^2\bigr]\\
&=   1-(\mathbf{b}\cdot\mathbf{c})^2
+\no{\bob-\boc(\inpr\bob\boc)}^2 =2\no{\mathbf{b}\times\mathbf{c}}^2.
\end{align*}
These are state-independent, as expected.

With the choices $\boa=(1,0,0), \bob= (0,1,0), \boc =(\cos\alpha,\sin\alpha,0), \bor= (0,0,1)$ the above scenario is just the experiment studied and realized  by the Vienna group \cite{Erh12}. Then, in particular,  $\mathbf r\cdot\mathbf a=\mathbf r\cdot\mathbf b=\inpr\bor\boc=0$, so that both  state-dependent errors become zero:
$\Delta(C_\rho,A_\rho)=0$
and $\Delta(D_\rho,B_\rho)=0$. By contrast,
$\epno(A,\rho)=\epno(\sigma_x,\rho) = 2\sin\frac{\alpha}2$,
$\epno(B,\rho)=\epno(\sigma_y,\rho)=\sqrt2\cos\alpha$; these are   bad overestimates of 
the state dependent error and disturbance for most values of $\alpha$. 
Curiously, the experimenters do not report a comparison of the values obtained for the quantities $\epno$ with an actual estimation of the error in measuring observable $C$ as an approximation of $A$; this would be of particular interest as the target and estimator observables do not commute; yet in the given state, the two observables are indistinguishable, while $\epno$ does not recognize this. This discrepancy should show up in an error analysis.

Instead of using a projective measurement of a ``misaligned'' sharp observable $C$ as an approximator to $A$ one may construct an explicit measurement scheme $\h M$ as an approximate $A$-measurement. Such a strategy was followed in the Toronto experiment \cite{Roz12}, which we reconstruct next. We take the parameters as used in that experiment. Thus, we fix $\mathbf a=(0,0,1)=\bok$ and consider a measurement scheme
$\h M=(\C^2,\sigma_z,U,|\phi\rangle\langle\phi|)$, where $\sigma_z$ is the  pointer, the coupling $U$ is the CNOT gate 
(in the canonical basis of $\C^2\otimes\C^2$), 
$\phi= \alpha|0\rangle+\beta|1\rangle$, $\alpha,\beta\in\R$, $\alpha^2+\beta^2=1$, (again in the canonical basis). 
The measured observable $C$ is then an unsharp version of the observable $A=E^{\bok}$, the spectral measure of $\sigma_z$,
$$
C_\pm= \tfrac 12(\idty\pm (2\alpha^2-1)\sigma_z).
$$
The distortion exerted by $\h M$ on the observable $B=E^{\boi}$
($\bob= (1,0,0)=\boi$)  then results in an observable $D$, where
$$ 
D_\pm 
 =
 \tfrac 12(\idty\pm 2\alpha\beta\sigma_x).
$$ 
These observables can also be written in terms of Bloch vector parametrization for $\phi$ using 
$\bos=(\sin\theta\cos\varphi,\sin\theta\sin\varphi,\cos\theta)$:
\begin{align*}
C_\pm&=\tfrac12(\idty\pm\cos\theta\sigma_z)=\tfrac12(\idty\pm(\mathbf{s}\cdot\mathbf{k})\,\mathbf{k}\cdot\boldsymbol{\sigma}),
\\
D_\pm&=\tfrac12(\idty\pm\sin\theta\cos\phi\sigma_x)=\tfrac12(\idty\pm(\mathbf{s}\cdot\mathbf{i})\,\mathbf{i}\cdot\boldsymbol{\sigma}).
\end{align*}

The sequential joint observable $M_{k,\ell}=\stfm(k)^*(B_\ell)$ [here $\stfm(k)$ denotes the conditional output channel associated with the
outcome $k$ of $\h M$, and $\stfm(k)^*$ its dual channel] thus realizes an approximate joint measurements of  $A=E^{\bok}$ and $B=E^{\boi}$.
This gives the following expressions for the state-dependent and maximized errors:
\begin{align*}
\Delta(C_\rho,A_\rho)^2&=|\mathbf{r}\cdot\mathbf{k}|\,\no{\mathbf{s}-\mathbf{k}}^2\le \no{\mathbf{s}-\mathbf{k}}^2=
\Delta(C,A)^2,\\
\Delta(D_\rho,B_\rho)^2&=|\mathbf{r}\cdot\mathbf{i}|\,\no{\mathbf{s}-\mathbf{i}}^2\le 
\no{\mathbf{s}-\mathbf{i}}^2=
\Delta(D,B)^2.
\end{align*}
If the initial state of the system is $\rho=\rho_\boj$, with $\mathbf r=\mathbf j$, then, again, both state-dependent errors vanish.

By contrast, the $\epno$ quantities are again state-independent and coincide, in fact, with the $\Delta$-errors:
\begin{align*}
\epno(\sigma_z,\rho)^2
&=\langle(1-\mathbf{s}\cdot\mathbf{k})^2\sigma_x^2\rangle_\rho+1-(\mathbf{s}\cdot\mathbf{k})^2=\no{\mathbf{s}-\mathbf{k}}^2,
\\
\epno(\sigma_x,\rho)^2
&=\langle(1-\mathbf{s}\cdot\mathbf{i})^2\sigma_z^2\rangle_\rho+1-(\mathbf{s}\cdot\mathbf{i})^2=\no{\mathbf{s}-\mathbf{i}}^2,
\end{align*}
again badly overestimating the state dependent errors.

The uncertainty relation for the maximized errors becomes here:
\begin{equation}\label{qubit-EDR2}
\begin{split}
\Delta(C,A)^2\,&+\,\Delta(D,B)^2
= \no{\mathbf{s}-\mathbf{k}}^2+\no{\mathbf{s}-\mathbf{i}}^2\\
&\ge2(2-\sqrt2).
\end{split}
\end{equation}
This is the optimal lower bound of \eqref{QUR}; it is reached with $\varphi=0$ and $\theta=\pi/4$, hence $\bos=(\boi+\bok)/\sqrt2$.

In the actual experiment \cite{Roz12} the numbers $\epno$ are determined using the weak measurement strategy suggested by \cite{LundWiseman2010}, thus confirming rather indirectly the quantum predictions for the expectations of second moments of the relevant difference observables. Again, no error analysis is reported  in \cite{Roz12} to check whether the $\epno$ numbers in question reflect the actual measurement errors.

In any case, the data that have been obtained in these experiments or could be obtained in variations of them 
can easily be used to test the error trade-off inequality \eqref{QUR} since the $\Delta$ errors are here found to be directly related to the corresponding $\epno$ numbers.

\section{Conclusion}

With the inequality \eqref{QUR} we have provided a general error trade-off relation for joint measurements of qubit observables in the spirit of Heisenberg's ideas of 1927. The additive form of this inequality can be matched with an additive form of preparation uncertainty relation, with a {\em state-independent} lower bound that only vanishes when the observables commute. We have also exhibited the true operational meaning of the quantities, $\epno$, in the qubit context, which were taken to represent error and disturbance in these experiments. Our analysis shows that Ozawa's inequality does not admit an interpretation as a trade-off between  error and disturbance for individual states. Rather than leading to a violation of a Heisenberg bound, the $\epno$ quantities were found themselves to obey Heisenberg-type trade-off relations. Finally we have identified possible tests of our new error relation that could be performed using the Vienna and Toronto experiments. We emphasize that such tests are not complete by simply measuring the $\epno$ or $\Delta$ quantities: a genuine test of error-error or error-disturbance trade-off relations must compare these data with an error analysis carried out for the joint measurements of $C$ and $D$ as approximations of $A$ and $B$, respectively, as indicated in figures 1 and 2.

\section*{Acknowledgements.}
This work is partly supported (P.B., P.L.) by the Academy of Finland, project no 138135, and EU COST Action MP1006. R.F.W.\ acknowledges support from the European network SIQS.

\section*{Appendix A. Calculation of Wasserstein distance}

We consider a slightly more general problem, that of minimizing the quantity
\[
\Delta^\gamma(E_\rho,F_\rho)^2=\iint(x-y)^2d\gamma(x,y)
\]
when $E$ has values $\pm 1$ and $F$ has values $a_+,a_-$, where we assume $a_+>a_-$. Our use of this will be to consider $F$ as an approximation to $E$. The Wasserstein distance should in fact vanish when the probabilities of $E$ and $F$ coincide for their corresponding values, $\pm 1\leftrightarrow a_\pm$. 

A general coupling is given by four positive numbers, 
\begin{align*}
(1,a_+)&\mapsto\gamma_{++}\equiv\gamma,\\
(1,a_-)&\mapsto\gamma_{+-}=E_\rho(+1)-\gamma,\\
(-1,a_+)&\mapsto\gamma_{-+}=F_\rho(a_+)-\gamma,\\
(-1,a_-)&\mapsto\gamma_{--}=1-E_\rho(+1)-F_\rho(a_+)+\gamma.
\end{align*}
It is then straightforward to obtain
\begin{align*}
{\Delta^\gamma}(E_\rho,F_\rho)^2&=(1+a_-)^2-4\gamma(a_+-a_-)-4E_\rho(+1)a_-\\
&\quad+F_\rho(a_+)\bigl[ (1+a_+)^2-(1+a_-)^2 \bigr]
\end{align*}
In order to minimise this quantity, $\gamma$ must be chosen as large as allowed by the positivity constraints (given that $a_+-a_->0$),
hence $\gamma=\min\{E_\rho(+1),F_\rho(a_+)\}$. Now it is easy to see that the minimum, $\Delta(E_\rho,F_\rho)$, can only vanish for $E_\rho(+1)=F_\rho(a_+)$ if $a_+=1$ and $a_-=-1$. In this case one obtains
\[
\Delta\bigr(E_\rho,F_\rho\bigr)^2=4|E_\rho(+1)-F_\rho(+1)|=2\bigl |e_0-f_0+\inpr\bor{(\boe-\bof)}\bigr|.
\]

\section*{Appendix B.  Proof sketch for the error trade-off inequality (1)}

This inequality is a direct translation, here for the $\Delta$-measure, of an  equivalent form proven in \cite{BuHe08} for the $D$-measure, using the proportionality of $\Delta^2$ with $D$. We sketch the steps of its derivation.
One first makes use of the reduction 
of \eqref{QUR} to the case where $c_0=1=d_0$. If $C=\frac 12\bigl[c_0\idty +\inpr\boc\bosig\bigr]$ and  
$D_+=\frac12\bigl[d_0\idty +\inpr\bod\bosig\bigr]$ are jointly measurable, then so are $C',D'$ with $C'_+=\frac12\bigl[(2-c_0)\idty +\inpr\boc\bosig\bigr]$ and  
$D'_+=\frac12\bigl[(2-d_0)\idty +\inpr\bod\bosig\bigr]$. It follows that the convex combinations of these observables are also jointly measurable \cite{BuHe08},
in particular $\tic$ and $\tid$ with
$\tic_+=\frac12(C_++C'_+)=\frac12(\idty+\inpr\boc\bosig)$ and $\tid_+=\frac12(D_++D'_+)=\frac12(\idty+\inpr\bod\bosig)$.
In addition we have that the errors do not increase:
\begin{align*}
\Delta\bigr(C,A\bigr)^2&\ge 2\no{\boc-\boa}=\Delta\bigr(\tic,A\bigr)^2 ,\\
\Delta\bigr(D,B\bigr)^2&\ge2\no{\bod-\bob}=\Delta\bigr(\tid,B\bigr)^2.
\end{align*}
This process of averaging can be understood as the transition to observables that are covariant under the shift group $\pm 1\mapsto \mp 1$ acting on the 
set $\{-1,+1\}$ \cite{BuHe08}. This group acts on $\tic$  and $\tid$ via the unitary operator \hbox{$U=\inpr{\bou}{\bosig}$,} with $\bou$ a unit vector perpendicular to $\boc$ and $\bod$, so that the covariance $U\tic_\pm U^*=\tic_\mp$ and $U\tid_\pm U^*=\tid_\mp$ holds.  We may therefore refer to the observables $\tic,\tid$ as {\em covariant}.
The compatibility of these covariant observables is equivalent to inequality \eqref{jm1}.

A similar convexity argument shows that if $\boc,\bod$ are not already in the plane spanned by $\boa,\bob$, then their projections into that plane define new observables which are again compatible and no worse approximations to $A,B$ than $C,D$. Hence we can assume that $\boc,\bod$ are in the plane spanned by $\boa, \bob$.

\begin{figure}[ht]
\centering
  \includegraphics[width=7cm]{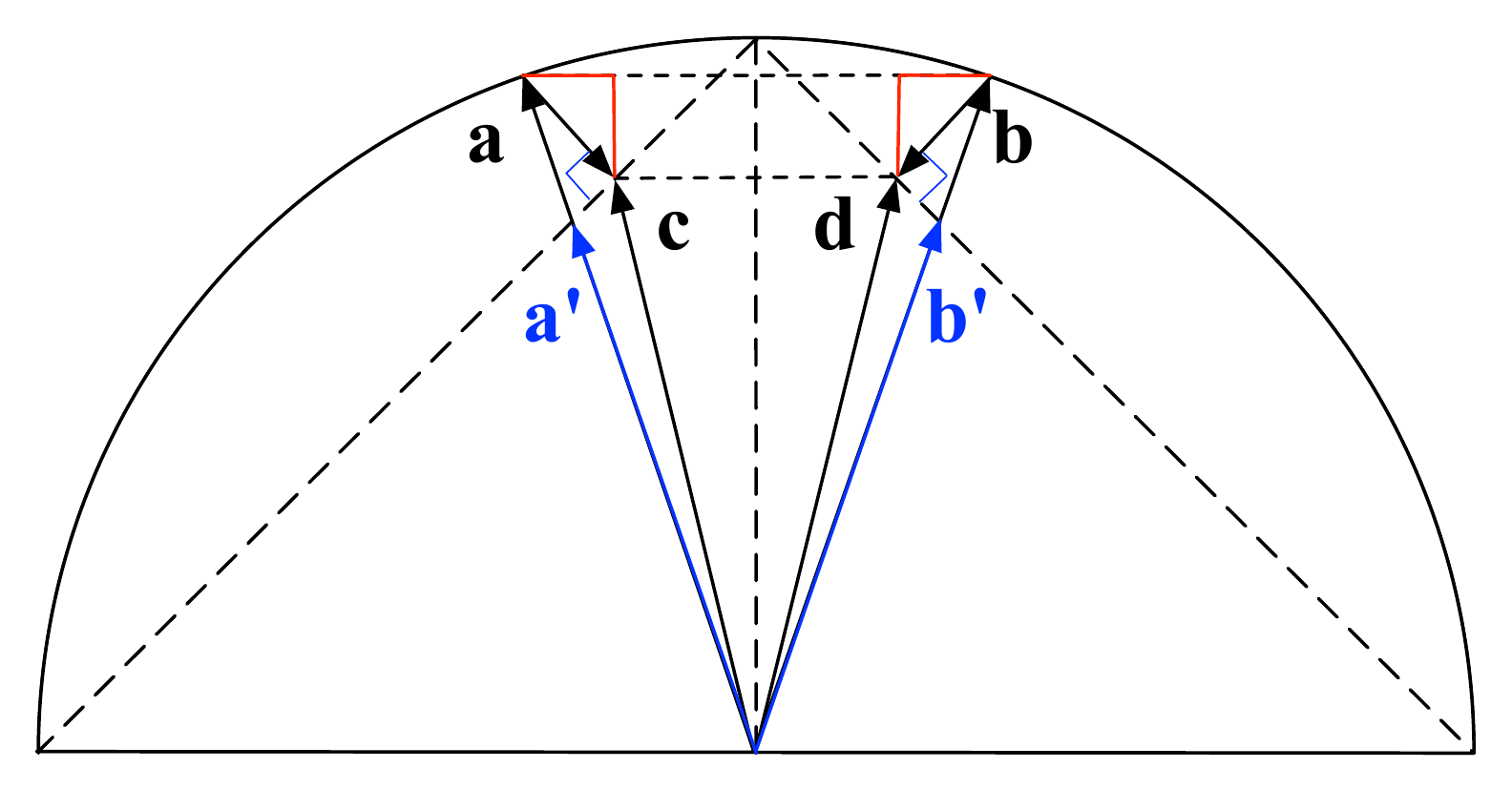}
	\caption{Optimal compatible approximations of sharp observables $A,B$ by covariant unsharp observables $C,D$. The compatibility of the
optimal pair $C,D$ has $\boc,\bod$  located on the dotted lines and the vectors $\boc-\boa$ and $\bod-\bob$ orthogonal to these dotted lines.
Vectors $\boa'$, $\bob'$ represent the best compatible approximators among the  ``smeared'' versions of $A,B$.}
		\label{Fig:compat}
\end{figure}

A simple geometric consideration shows that the minimum of the left hand side of \eqref{QUR} must be attained for approximators $C,D$ whose vectors $\boc,\bod$ have equal length and are located symmetrically relative to $\boa,\bob$, as shown in figure \ref{Fig:compat}.  (In fact, using once more the preservation of compatibility under convex mixings of observables, it is straightforward to see that any asymmetric constellation of vectors $\boc,\bod$ can be transformed into a symmetric one for which the errors are not greater.) Analysis of the right-angled triangle with vertices given by the end points of $\boa$, $\boc$ and the intersection between the vertical line through $\boc$ and the horizontal line connecting $\boa$ and $\bob$ (and similarly on the side of $\bob$) immediately gives the relations $\frac12\bigl[ \no{\boa-\bob}-\no{\boc-\bod}\bigr]=\frac12\bigl[ \no{\boa+\bob}-\no{\boc+\bod}\bigr]=\no{\boc-\boa}/\sqrt2=\no{\bod-\bob}/\sqrt2$; hence the lower bound in \eqref{QUR} follows via the compatibility constraint $\no{\boc-\bod}+\no{\boc+\bod}=2$.

\section*{Appendix C. Compatibility criterion and joint observable}\label{app:compat}

Compatible observables $C,D$ with $C_\pm=\frac 12(\idty\pm\inpr\boc\bosig)$, $D_\pm=\frac12(\idty\pm\inpr\bod\bosig)$
arise as marginals of the operator measure $G:\,k,\ell\mapsto G_{k\ell}$, $k,\ell=\pm 1$, where
\begin{align*}
G_{+,\pm}&=\tfrac14\left(1\pm\inpr\boc\bod\right)\idty+\tfrac14\left(\boc\pm\bod\right)\cdot\bosig,\\
G_{-,\pm}&=\tfrac14\left(1\mp\inpr\boc\bod\right)\idty-\tfrac14\left(\boc\mp\bod\right)\cdot\bosig,
\end{align*}
Note the marginality relation $C_\pm=G_{\pm,+}+G_{\pm,-}$ and $D_\pm=G_{+,\pm}+G_{-,\pm}$. For $G$ to be an observable,
the operators $G_{k\ell}$ must be positive, that is, $1\pm\inpr\boc\bod\ge\no{\boc\pm\bod}$. This implies immediately, and is in fact equivalent to, \eqref{jm1}. (Equivalence follows easily via \eqref{jm2}.) The proof of the necessity of \eqref{jm1} for the compatibility of $C,D$ is slightly more involved \cite{Busch86}.

%\bibliography{UR}


\end{document}